\documentclass[fleqn,10pt]{wlscirep}
\usepackage{bbding}
\usepackage{amsthm}
\usepackage{dcolumn}
\usepackage{bm}
\usepackage{subfigure}
\usepackage{epsfig,graphicx,times}
\usepackage{amstext}
\usepackage{amsmath}
\usepackage{amssymb}
\usepackage{graphicx}
\begin{document}
\title{Optomechanically induced transparency in multi-cavity optomechanical system with and without one two-level atom}
\author[1]{Amjad Sohail}
\author[1]{Yang Zhang}
\author[1]{Jun Zhang}
\author[1,*]{Chang-shui Yu}
\affil[1]{School of Physics and Optoelectronic Technology, Dalian University of
Technology, Dalian 116024, P.R. China }

\affil[*]{quaninformation@sina.com; ycs@dlut.edu.cn}


\begin{abstract}
We analytically study the optomechanically induced transparency (OMIT)  in the $N$-cavity system with the \textit{N}th cavity driven by pump, probing laser
fields and the \textit{1}st cavity coupled to mechanical oscillator. We also consider that one atom could be trapped in the \textit{i}th cavity. Instead of only illustrating the OMIT in such a system, we are
interested in how the number of OMIT windows is influenced by the cavities
and the atom and what roles the atom could play  in different cavities. In the resolved sideband  regime, we find that, the number of
cavities precisely determines the maximal number of OMIT windows. It is interesting that, when the two-level atom is trapped in the even-labeled cavity,  the central absorptive peak (odd $N$) or dip (even $N$) is split and forms an extra OMIT window, but if the atom is trapped in the
odd-labeled cavity,  the central absorptive peak (odd $N$)  or dip  (even $N$) is only broadened and thus changes  the width of the OMIT windows rather than induces an extra window.\end{abstract}

\flushbottom
\maketitle

\thispagestyle{empty}

\section*{Introduction}

Cavity optomechanical system (OMS) has recently attracted increasing interest in both theory and experiment (Ref. \cite{Aspelmeyer} and references therein). It usually composed of two mirrors with
one fixed and the other movable or a micro-mechanical membrane
oscillating inside two fixed mirrors. Such a system demonstrates the interaction between the movable oscillator and the optical field in the cavity
via the radiation pressure and becomes a platform for the study  \cite{e1,e2,e3, e4,AN, Rabl, Purdy,Tian,Chang, Akram, Clerk, Karuza, Hammerer, Saif, Islam} of  quantum
ground-state cooling \cite{c1,TJD,c3,Meystre} , strong coupling dynamics \cite{Aspelmeyer, Connell, Kiesel, Rabl} and other coherent dynamics in
microscopic and macroscopic domains \cite{Rogers, Kippenberg, Restrepo} .
When a strong laser field drives the optomechanical
cavity, an analogue of electromagnetically induced transparency (EIT) for the output at the frequency of the weak
detecting field could appear \cite{Agarwal, SM} . Such an EIT-like phenomenon is usually
called as the optomechanically induced transparency (OMIT) which is equivalent to the case of
two coupled harmonic oscillators \cite{Connell} and has been demonstrated in
experiments \cite{t2,Safavi,t4} .
OMIT has also been  widely investigated in diverse aspects including  the cases with higher-order sidebands \cite {HX} or in the nonlinear regime \cite{MAL,KB,AK}, OMIT in the cavity with membranes \cite{MK,t6} and so on. In particular, OMIT has shown many potential applications  in control of light speed \cite{Safavi}, charge measurement \cite{Char},  single photon router \cite{router} and so on, which forms  the further motivations to study OMIT.

Introducing the atomic freedom into OMS can not only strengthen the coupling but also allow  rich physics via enhanced nonlinearities \cite{at1,KQu,HW}. It has been applied to improve optomechanical cooling \cite{atc1,atc2,atc3} and even the ground-state cooling outside the resolved sideband regime \cite{ures}. In particular, it is shown \cite{Ian}  that a two-level atomic
ensemble coupled to OMS can both enhance the photon-phonon coupling through radiation pressure and
broaden the transparency windows. In addition, coupled-cavity array related to the 1D waveguide or atoms has been widely studied in the control of photon transport such as quantum router \cite{LN,CHY,LZ,JL}. Does the multiple-cavity quantum optomechanics bring new insight into OMIT? How can the OMIT be controlled if introducing the atomic freedom into the multiple-cavity system?

In this paper, we address the above questions by investigating the OMIT phenomenon in multiple-cavity optomechanical
system coupled to one two-level atom.  \textit{Here instead of only illustrating the OMIT in such a system, we are especially
interested in how the number of OMIT windows  is
related to the number of the cavities as well as the potential trapped atom and what roles the atom could play in different cavities.}
Through our analytic calculations, it is shown that the maximal number of
OMIT windows is precisely determined by the cavity number, if there does not
exist any atom in the multi-cavity system. In particular, we find that the
atom trapped in different cavities will play different roles in OMIT. When
one atom is trapped in even-labeled cavity, the central absorptive peak (odd $N$) or dip (even $N$) is split and forms an extra OMIT window, but when 
the atom is trapped in odd-labeled cavity, the central absorptive peak (odd $N$) or dip (even $N$) is only broadened and thus  changes the width of
the OMIT windows instead of inducing the extra window.  In addition, we also find that the multiple OMIT windows are the result of the coupling of multiple cavities irrespective of the participation of the mechanical oscillator. A numerical simulation is also given to support our results.

\section*{Results}
\textbf{The model.} The optomechanical system under consideration is shown schematically in Fig. 1.  The system includes $N$ cavities labelled by $%
1,2,\cdots, N$ with the frequency of $j$th cavity denoted by $\omega_j$. The $n$th and $(n+1)$th cavities with $n\neq N$ are connected through
tunneling parameters (hopping rates) $g_{n}$.  Such a coupled cavity array (2D) has been systematically studied in various cases in Ref. \cite{Plenior} and later considered in the single-photon router \cite{LZ,JL}. Here we only consider 1D cavity chain, in particular, we let one end mirror of Cavity $1$ be movable as shown in Fig. 1. Thus it forms an optomechanical system. Cavity $N$ is separately driven by one coupling field $%
\varepsilon _{c}$ and one probing field $\varepsilon _{p}$.  In addition, we
assume that one two-level atom could
be trapped in the $i$th cavity $1\leq i\leq N$ with $g_{a}$ denoting the atom-cavity coupling strength.   In this model, the
optical modes are described by annihilation (creation) operators $%
c_{n}(c_{n}^{\dagger })$ and the mechanical mode is represented by $%
b(b^{\dagger })$ which is equivalent to the description by $x_m$ and $p_m$. This similar description can be found in Ref. \cite{HW}. Let the frequency of the coupling field be $\omega
_{c}$, so in the rotating frame at $\omega
_{c}$, the Hamiltonian of our system reads
\begin{eqnarray}
H &=&\sum_{j}\Delta _{j}c_{j}^{\dagger }c_{j}+\omega _{m}b^{\dag }b+\frac{1}{%
2}\Delta _{a}\sigma _{z}+i\varepsilon _{c}\left( c_{N}^{\dagger }-c_{N}\right) +i\varepsilon
_{p}\left( c_{N}^{\dagger }e^{-i\Delta t}-c_{N}e^{i\Delta t}\right)  \notag
\\
&&+g_{a}\left( c_{i}\sigma _{+}+c_{i}^{\dagger }\sigma _{-}\right)
-gc_{1}^{\dagger }c_{1}\left( b^{\dag }+b\right)+\sum\limits_{n=1}^{N-1}g_{n}\left( c_{n+1}^{\dagger }c_{n}+c_{n}^{\dagger
}c_{n+1}\right) \label{Ham}
\end{eqnarray}
with $\omega _{p}$,  $\omega _{a}$ representing the frequency of the probing field and the atomic transition frequency. In  Eq. (\ref{Ham}) the first three terms, respectively, denote the free Hamiltonian  for the cavities, the movable mirror and the trapped atom with $\Delta _{j}=\omega _{j}-\omega _{c}$  $\Delta =\omega _{p}-\omega
_{c} $ and $\Delta _{a}=\omega _{a}-\omega _{c}$, the last two terms in first line corresponds to the interaction of the \textit{N}th cavity driven by the coupling field $\varepsilon_c$ and the probing field $\varepsilon_p$. The first term in the second line of Eq. (\ref{Ham}) describes the interaction between the atom and the \textit{i}th cavity,  the second term corresponds to the interaction between the \textit{1}th cavity and the movable mirror via the radiation pressure, and the last term describes the hopping between the two adjacent cavities.  In addition, $g$ in Eq. (\ref{Ham}) denotes
the coupling strength between the $1st$ cavity and the mechanical
oscillator. It is obvious that $\Delta _{a}=g_{a}=0$ means no atom in the
cavities.

\textbf{The dynamics.} Based on the above Hamiltonian, one can easily obtain the Langevin
Equations for the operators. So the corresponding equations for the mean
value of operators in the mean-field approximation, viz, $\left\langle
st\right\rangle =\left\langle s\right\rangle \left\langle t\right\rangle $,
can be given by

\begin{figure}[tbp]
\centering
\includegraphics[width=1\columnwidth,height=1.25in]{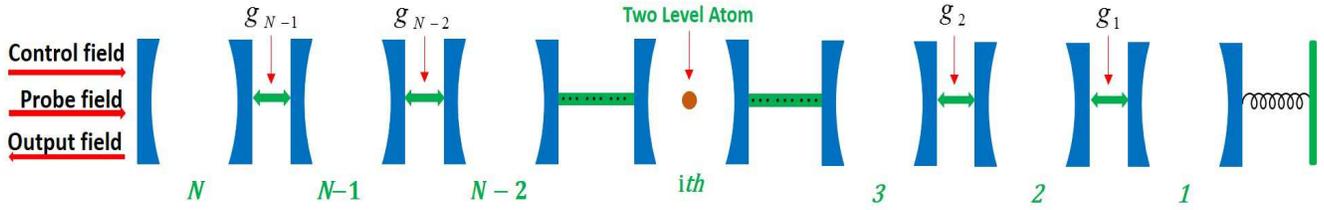} \centering
\caption{(color online).Schematic diagram of $N$ cavities connected through
tunneling parameters $g_{n}$. A strong driving field and a weak probing field
are injected into cavity $N$ while the first cavity is coupled with
mechanical resonator}
\end{figure}

\begin{equation}
\left\langle \dot{c}_{N}\right\rangle =-\left( \kappa _{N}+i\Delta
_{N}\right) \left\langle c_{N}\right\rangle -ig_{N-1}\left\langle
c_{N-1}\right\rangle +\varepsilon _{c}+\varepsilon _{p}e^{-i\Delta t},
\label{(2)}
\end{equation}%
\begin{equation}
\left\langle \dot{c}_{n}\right\rangle =-\left( \kappa _{n}+i\Delta
_{n}\right) \left\langle c_{n}\right\rangle -i\left( g_{n-1}\left\langle
c_{n-1}\right\rangle +g_{n}\left\langle c_{n+1}\right\rangle \right) ,n\neq
1,i,N, \label{(3)}
\end{equation}%
\begin{equation}
\left\langle \dot{c}_{i}\right\rangle =-\left( \kappa _{i}+i\Delta
_{i}\right) \left\langle c_{i}\right\rangle -ig_{a}\left\langle \sigma
_{-}\right\rangle -i\left( g_{i-1}\left\langle c_{i-1}\right\rangle
+g_{i}\left\langle c_{i+1}\right\rangle \right) ,~i\neq 1,N, \label{(4)}
\end{equation}%
\begin{equation}
\left\langle \dot{c}_{1}\right\rangle =-\left( \kappa _{1}+i\tilde{\Delta}_{1}\right) \left\langle c_{1}\right\rangle -ig_{1}\left\langle
c_{2}\right\rangle +ig\left\langle c_{1}\right\rangle \left( \left\langle
b^{\dag }\right\rangle +\left\langle b\right\rangle \right) , \label{(5)}
\end{equation}%
\begin{equation}
\left\langle \dot{b}\right\rangle =-\left( \gamma _{m}+i\omega _{m}\right)
\left\langle b\right\rangle +ig\left\vert \left\langle c_{1}\right\rangle
\right\vert ^{2},  \label{(6)}
\end{equation}%
\begin{equation}
\left\langle \dot{\sigma}_{-}\right\rangle =-\left( \gamma _{a}+i\Delta
_{a}\right) \left\langle \sigma _{-}\right\rangle +ig_{a}\left\langle
c_{i}\right\rangle \left\langle \sigma _{z}\right\rangle,  \label{(7)}
\end{equation}%
\begin{equation}
\left\langle \dot{\sigma}_{z}\right\rangle =-2\left( 1+\left\langle \sigma
_{z}\right\rangle \right) \gamma _{a}+2ig_{a}\left( \left\langle c_{i}^{\dag
}\right\rangle \left\langle \sigma _{-}\right\rangle +\left\langle
c_{i}\right\rangle \left\langle \sigma _{+}\right\rangle \right).  \label{(8)}
\end{equation}%
Here $\kappa _{n}$ denotes the leakage of \textit{n}th cavity and $\gamma
_{m}$ and $\gamma _{a}$, respectively, represent the decay rates of the
mechanical oscillator and the atom. If the atom is trapped in the first
cavity, Eq. (\ref{(5)}) should be replaced by
\begin{eqnarray}
\left\langle \dot{c}_{1}\right\rangle &=&-\left( \kappa _{1}+i\tilde{\Delta}_{1}\right) \left\langle c_{1}\right\rangle -ig_{1}\left\langle
c_{2}\right\rangle-ig_{a}\left\langle \sigma _{-}\right\rangle +ig\left\langle
c_{1}\right\rangle \left( \left\langle b^{\dag }\right\rangle +\left\langle
b\right\rangle \right).
\end{eqnarray}%
If the atom is trapped in $N$th cavity, Eq. (\ref{(2)}) should be replaced by
\begin{eqnarray}
\left\langle \dot{c}_{N}\right\rangle &=&-\left( \kappa _{N}+i\Delta
_{N}\right) \left\langle c_{N}\right\rangle -ig_{N-1}\left\langle
c_{N-1}\right\rangle-ig_{a}\left\langle \sigma _{-}\right\rangle +\varepsilon _{c}+\varepsilon
_{p}e^{-i\Delta t}.\label{4}
\end{eqnarray}
In order to solve the dynamics, we suppose
\begin{equation}
\left\langle \hat{O}(t)\right\rangle =\bar{O}+\delta {O}(t)=\bar{O}%
+O_{-}e^{-i\Delta t}+O_{+}e^{i\Delta t},  \label{O}
\end{equation}%
for any operator $\hat{O}(t)$ with $\bar{O}$ denoting the steady-state value
without $\varepsilon _{p}$ and $\delta {O}=O_{-}e^{-i\Delta
t}+O_{+}e^{i\Delta t}$ induced by the weak probing field. Substituting Eq. (\ref{O}) into Eqs. (\ref{(2)}-\ref{4}), one can obtain  an equation array for $\bar{O}$ which has the same form as Eqs. (\ref{(2)}-\ref{4}) except $\varepsilon_p=0$ and $\dot{\bar{O}}=0$. This equation arrays are omitted here.  In addition, one can also obtain an equation array for $\delta O(t)$ which is given in the Methods (Eq. (\ref{(17)}-\ref{(22)})).
By solving the equations for $\bar{O}$, one can find that
 \begin{equation}\bar{\lambda}=\bar{b}+\bar{b}%
^{\ast }=\frac{2\omega _{m}g}{\omega _{m}^{2}+\gamma _{m}^{2}}\left\vert
\bar{c}_{1}\right\vert ^{2},\label{1ping}\end{equation}
and
\begin{equation}\bar{\sigma}_{z}=\frac{\gamma _{a}\left(
\gamma _{a}^{2}+\Delta _{a}^{2}\right) }{2i\Delta _{a}g_{a}^{2}\left\vert
c_{i}\right\vert ^{2}-\gamma _{a}\left( \gamma _{a}^{2}+\Delta
_{a}^{2}\right) }.\label{2ping}\end{equation}
In addition, considering Eq. (\ref{O}), one can easily find the equations for $O_{\pm}$. However, for the purpose of this paper, we only provide the equations for $O_{-}$ in the Methods Eq. (\ref{(23)}-\ref{(28)}) within the resolved sideband regime, i.e., $\omega _{m}\gg \kappa $ and  $%
\Delta _{n}=\tilde{\Delta}_{1}=\Delta _{a}=\omega _{m}$ where $\tilde{%
\Delta}_{1}=\Delta _{1}-g\bar{\lambda}$.
These equations provide the fundamental description of the dynamics of the model considered here.

\textbf{Output field.} In order to reveal the OMIT, we will have to find out the response of the
system to the probing frequency, which can be detected by the output field.
Based on the input--output theory \cite{Optics} , we can obtain
\begin{equation}
\varepsilon _{out,p}e^{-i\Delta t}+\varepsilon _{p}e^{-i\Delta t}+\varepsilon
_{c}=2\kappa _{N}\left\langle c_{N}\right\rangle .  \label{7}
\end{equation}%
Substituting Eq. (\ref{O}) into Eq. (\ref{7}), one can find that the total
output field at the probing frequency $\omega _{p}$ can be given by%
\begin{equation}
\varepsilon _{T}=\frac{\varepsilon _{out,p}}{\varepsilon _{p}}+1=\frac{%
2\kappa _{N}c_{N,-}}{\varepsilon _{p}}=\chi _{p}+i\tilde{\chi}_{p}.
\end{equation}%
It is clear that $\chi _{p}=$Re$\left( \varepsilon _{T}\right) $ and $%
\widetilde{\chi }_{p}=$Im$\left( \varepsilon _{T}\right) $ are  the
in-phase and out-of-phase quadratures of the output probing field,
representing the absorptive and dispersive behavior of the output probing
field, respectively. The quadrature can be measured via the homodyne
technique \cite{Optics} . So the next task is to find $c_{N,-}$.  In order to
gain more physical insight, we only consider the system in the sideband resolved regime.
Thus $c_{N,-}$ can be easily obtained by solving Eq. (\ref{(23)}-\ref{(28)}). So the output field $\varepsilon_T$ can be directly given by
\begin{equation}
\varepsilon _{T}=2\kappa _{N}c_{N,-}=\frac{2\kappa _{N}}{\left( \kappa
_{N}-ix\right) +\frac{g_{N-1}^{2}}{\kappa _{N-1}-ix+\frac{g_{N-2}^{2}}{%
\kappa _{N-2}-ix+\frac{g_{N-3}^{2}}{%
\begin{array}{c}
\ddots  \\
\kappa _{i}-ix+\frac{g_{a}^{2}\left\vert \bar{\sigma}_{z}\right\vert ^{2}}{%
\gamma _{a}-ix}+\frac{g_{i-1}^{2}}{\kappa _{i-1}-ix+\frac{g_{i-2}^{2}}{%
\begin{array}{c}
\ddots  \\
\kappa _{2}-ix+\frac{g_{1}^{2}}{\kappa _{1}-ix+\frac{\left\vert G\right\vert
^{2}}{\gamma _{m}-ix}}%
\end{array}%
}}%
\end{array}%
}}}},\label{result}
\end{equation}%
where $x=\Delta -\omega _{m}$ and $G=g\bar{c}_{1}$ is the effective optomechanical rate.
In above equation, the first line of the denominator
represents two cavities with radiative decays $\kappa _{N}$ and $\kappa
_{N-1}$ are connected through their coupling strength $g_{N-1}$. Second line
represents two cavities with radiative decays $\kappa _{N-1}$ and $\kappa
_{N-2}$ are connected through their coupling strength $g_{N-2}$ and so on.
The \textit{1}st cavity in the last line is coupled to the mechanical oscillator by
an effective coupling $G$. In addition, an extra term $\frac{%
g_{a}^{2}\left\vert \bar{\sigma}_{z}\right\vert ^{2}}{\gamma _{a}-ix}$
corresponding to the atomic contribution appears in the $\kappa _{i}$ line
which denotes the atom is coupled to the $i$th cavity with an effective
coupling $g_{a}\left\vert \bar{\sigma}_{z}\right\vert $. Certainly, if the
atom is trapped in the first cavity, this term will appear in the last line.
If the atom is placed in the $N$th cavity, it will appear in the first line
of the denominator. It is obvious that the output field depends on both the
parameters of the system and the steady-state values of $c_{1}$ and $c_{i}$.
These two values can be determined by solving the equations for all $\bar{O}$
which have been omitted here. But the concrete expressions of $c_{1}$ and $%
c_{i}$ are quite complicated, so it is impossible to present the concrete
forms. It is fortunate that this does not influence our understanding on the
OMIT window numbers.  One can find from the latter part that the values of $%
c_{1}$ and $c_{i}$ only affect the width of the OMIT windows. In this sense, it doesn't matter whether they can be simultaneously assigned by some values. Therefore, for
simplicity, one can select that $\bar{\sigma}_{z}=-1$ and $G$ can be given by any reasonable and convenient assignment.

\textbf{OMIT windows.} The OMIT is signaled by the simultaneously vanishing absorption and
dispersion, which is further related to the simultaneously vanishing $\chi_p$ and $\tilde{\chi}_p$, that is $\varepsilon_T$. In order to show the OMIT windows as many as possible, we restrict ourselves to the weak dissipative regime, i.e., $g_i\gtrsim\kappa_N\gg\kappa
_{i},\gamma_{m/a} $, to discuss the points where $\varepsilon _{T}$ vanishes. This is also supported by our latter numerical procedures.
\begin{figure}[tbp]
\centering
\includegraphics[width=1\columnwidth,height=2in]{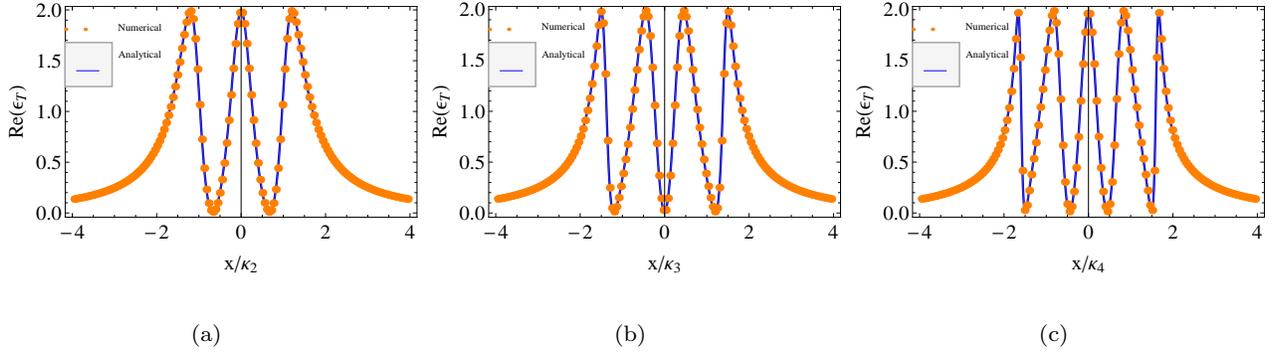}
\caption{(color online). The Real part Re$(\protect\varepsilon _{T})$ in the absence of
atom-field coupling, as a function of $x/\protect\kappa _{N}$ for (a) two
cavities, (b) three cavities and (c) four cavities with the parameters $%
\protect\omega _{m}=2\protect\pi \times 51.8MHz$, $\protect\gamma _{m}=2%
\protect\pi \times 41KHz$, $G=2\protect\pi \times 10Mz$, $\protect\kappa %
_{N}=2\protect\pi \times 15MHz$, $\protect\kappa _{1}=\protect\kappa _{2}=\protect\kappa _{3}=\cdots=2\protect\pi \times
0.027MHz$, and the coupling
rates $g_{1}=g_{2}=g_{3}=\protect\kappa _{N}$. The solid lines show the analytic expressions given by Eq. (16), but the dotted lines represent the solution numerically solved from Eq. (23-28), which guarantees the validity of our analytical result Eq. (16).}
\end{figure}
\begin{figure}[tbp]
\centering
\includegraphics[width=0.75\columnwidth,height=3in]{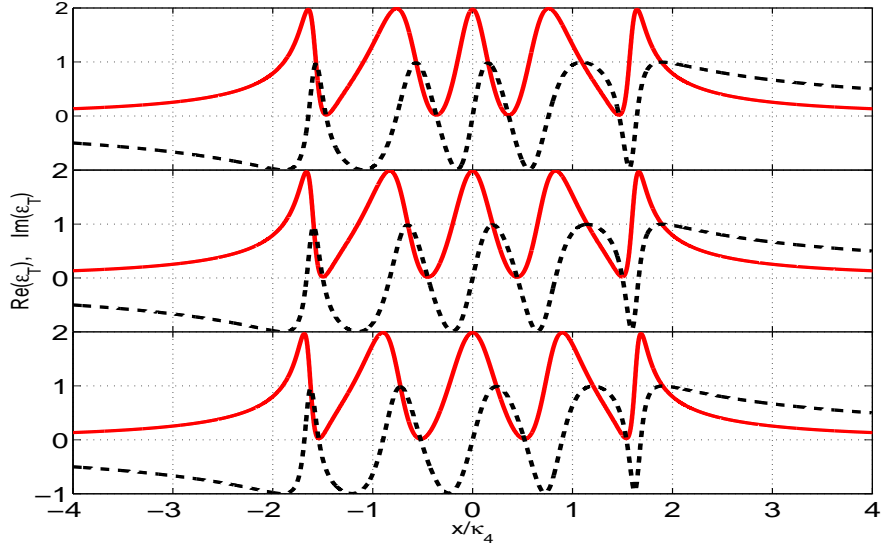} \centering
\caption{(color online). Real part Re($\protect\varepsilon _{T}$)(solid red) and the
imaginary part Im($\protect\varepsilon _{T}$)(dashed black) as a function of $x/\protect%
\kappa _{4}$ for four cavities. The three subplots from above to bottom,  respectively, correspond to $G=8MHz$, $G=10MHz$ and $G=12MHz$. The other parameters are the same as in Fig. $2$. One can see that the width of the central absorptive peak becomes wide with
the increasing of  $G$.}
\end{figure}%

\textit{Without atom.}-If there does not exist any atom in the
optomechanical system, the term with $g_{a}$ vanishes due to $g_{a}=0$. In
this case, the vanishing $\varepsilon _{T}$ means that the denominator
approaches infinity which can be further determined by the vanishing
denominator corresponding to the numerator $\left\vert g_{N-1}\right\vert
^{2}$. It is obvious that the condition with such a vanishing denominator
corresponds to an equation with $N$ degrees. Therefore it has at most $N$
different roots. This means that such an optomechanical system has at most $N$
OMIT windows. To give an intuitive illustration of the OMIT, we numerically
evaluate OMIT and demonstrate the multiple transparency windows due to the
interaction between cavity fields and the mechanical oscillator. We take the
parameters from \cite{Lin, Zheng,  Astafiev} where, the damping rate
of mechanical oscillator $\gamma _{m}=2\pi \times 41$ kHz, decay rate of the
driven cavity field $\kappa _{N}=2\pi \times 15$ MHz and the frequency of
oscillator $\omega _{m}=2\pi \times 51$ MHz. For the case of the resolved  sideband
regime, i.e. the mechanical frequency is much greater than the decays and $%
\Delta _{n}=\tilde{\Delta}_{1}=\Delta _{a}=\omega _{m}$, we plot the phase
quadratures of the output probing fields for a system with two, three and four
cavities in Fig. 2  which exhibits two windows, three windows and four
windows respectively. We assume that the $1st$ cavity coupled to mechanical
oscillator with $G=2\pi \times 12$ MHz. The multiple transparency windows
display that the optomechanical system becomes simultaneously transparent to
the probing field at multiple different frequencies, which is the
result of the destructive interferences between the input probing field and
the anti-Stokes fields generated by the interactions of the coupling field
with the multiple cavities. In addition, in order to show the effects of $G$, we plot Fig. 3 with different choices of $G$. One can find that the larger $G$ corresponds to the wider central absorptive peak (or dip for odd number of cavities) in the valid range of $G$. Numerical results show that the interval that the OMITs occur (from about $-2$ to $2$ in all the figures) is almost independent of the numbers of cavities. In fact, the width is determined by all the hopping rate $g_n$. Here in order to find out many enough OMIT windows, we let all $g_n=\kappa_N$, so the interval (if defined by the half width) is slightly changed. Under this condition, by numerical demonstrations, we find that the half width is increased with $N$. In particular, one can easily prove that when $N$ tends to infinity, the half width is just 4.  So when the central absorptive peak or dip gets wider, and the others  get narrower due to the fixed interval.  In one word, the value of $G$ only affects the width of the transparency window instead of
the maximal number of the OMIT window.

\textit{One atom in one cavity.}-Since we have set $g_i\gtrsim\kappa_N\gg\kappa
_{i},\gamma _{i}$, for an intuitive understanding of the number of OMIT
windows, one can safely neglect the dissipative constants which contributes
to the level width of the cavity as well as the atom. Under such a
condition, one can find that there exist two cases in our optomechanical
system. 1) \textbf{The atom is trapped in the odd-labeled cavity.} In this case, one
can see that the extra term $\frac{g_{a}^{2}\left\vert \bar{\sigma}%
_{z}\right\vert ^{2}}{\gamma _{a}-ix}$ can only exist in the lines
corresponding to $\kappa _{1},\kappa _{3},\cdots $. The contribution of such
an extra atomic term is mathematically to increase the numerator of the same
line and physically to directly broaden the central absorptive peak for even $N$ (or absorptive dip for odd $N$) and then to change the width of the OMIT windows, which is analogous to increasing $G$ in the  case without atom. The most obvious
example is when the atom is trapped in the first cavity. One can easily find
that for weak $\gamma _{a}$ and $\gamma _{m}$, the atomic term can be
approximately absorbed in the term corresponding to the mechanical
oscillator and the net result is equivalent to increasing $\left\vert
g_{1}\right\vert ^{2}$. 2) \textbf{The atom is trapped in the even-labeled cavity.}
In this case, the extra atomic term can lead to that the degree of the
equation of the vanishing denominator corresponding to the numerator $%
\left\vert g_{N-1}\right\vert ^{2}$ is added by $1$.  So when the atom is
trapped in the even-labeled cavity, one can find one more extra OMIT window
compared with the case without any atom. Similarly, in order to give an
illustration of these different cases, we numerically evaluate the OMIT. We
plot the figure in Fig. 4 with $g_{a}=2\pi \times 10$ MHz and $%
\gamma _{a}=2\pi \times 0.01$ MHz.  However, we don't plot the imaginary part Im($\protect\varepsilon _{T}$) for the sharp illustration.  We observe that, in four-cavity system,
the width of the central absorptive peak tends to become wide through embedding the
atom into the cavity $1$ or cavity $3$ as shown in Fig. 4 (a). But, when
the atom is placed in cavity $2$ or $4$, we have found the resonant
character of the weak probing field changes and the central absorptive peak splits. Hence four OMIT windows transfigure
to a penta OMIT window, as shown in Fig. 4 (c). Similarly, in Fig. 4 (b) and Fig. 4 (d) that correspond to the cases of three cavities, one 
can find that the atom will directly lead to the broadening or splitting of the central absorptive dips instead of absorptive peaks.

\textit{The role of the mechanical oscillator}.-Actually the physical
mechanism of the mechanical oscillator about the production of OMIT has been
well known \cite{Agarwal,SM} . In this part, we are only interested in
how the existence of the mechanical oscillator affects the number of OMIT
windows. If there does not exist any mechanical oscillator, that means $G=0$%
. If the atom is trapped in the first cavity under this condition, the number of the OMIT
windows will keep invariant, but the width of the OMIT window will become
narrow. This could be equivalently understood as the case without atom in the optomechanical system. That is, the role of the mechanical oscillator is to broaden the OMIT window in this case. In other cases, that is, no atom exists or the atom is only trapped
in the even-labeled cavity and so on, one can easily find that the OMIT
windows will be decreased by 1. In this case, one can draw the conclusion that
the mechanical oscillator contributes an OMIT
window. In this sense, we can say that the multiple OMIT windows should come
from the coupling of the multiple cavities instead of the direct
participation of the mechanical oscillator.
\begin{figure}[tbp]
\centering
\includegraphics[width=1\columnwidth,height=2.5in]{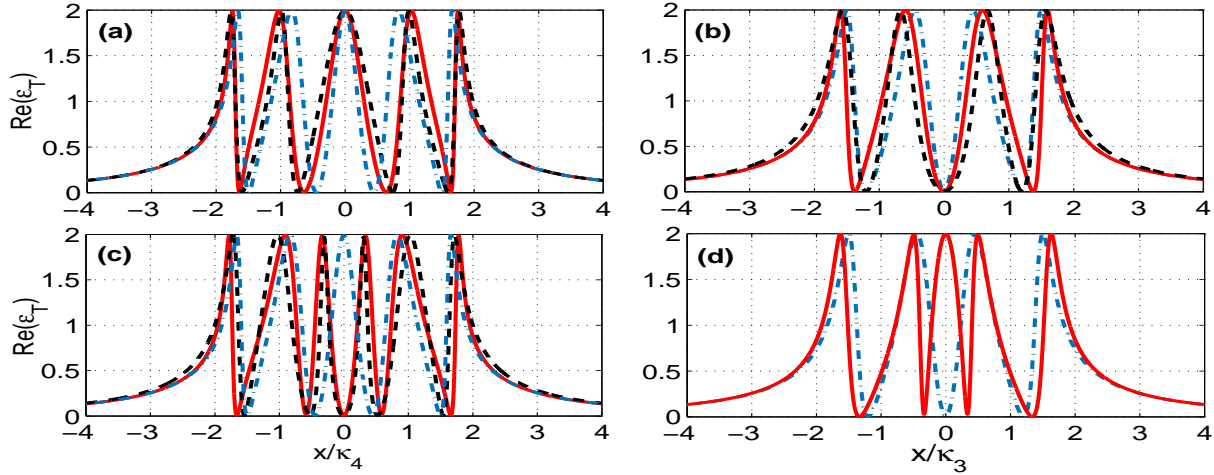} \centering
\caption{(color online). Real part Re$(\protect\varepsilon _{T})$ of the output field as a function of $x/\kappa_{3/4}$ with $\protect\gamma _{a}=2\protect\pi %
\times 0.01MHz$ and $g_{a}=2\protect\pi \times 10MHz$. The other parameters are the same in
Fig. $2$.  (a) and (c) illustrate cases with four cavities, where (a) corresponds to the atom  trapped in
cavity $1$ (solid red) and  cavity $3$ (dashed black) and (c) corresponds to the atom in cavity $2$ (solid red) and cavity $4$ (dashed black). (b) and (d) correspond to the cases with three cavities, where (b) illustates the atom trapped
 in cavity $1$ (solid red) and  cavity $3$ (dashed black) and (d) shows the atom trapped in cavity $2$ (solid red). The dashed blue lines in all the figures mean no trapped atom. }
\end{figure}
\section*{Discussions and Conclusion}
Before the end, we would like to emphasize that similar to multiple EIT windows, the multiple OMIT windows permit
the probing light with different frequencies to transmit simultaneously. So the OMIT with multiple windows could also be used
in multi-channel optical communication and multichannel
quantum information processing \cite{SMM}. OMIT is also closely related to the superluminal and ultraslow light
propagation \cite{TJD,Tar}, the quantum router \cite{router}, charge measurement\cite{Char} and so on. Hence, OMIT with multiple transparency windows 
could mean wider applications. 
In addition, the experimental realization of coupled cavity array is systematically reviewed in Ref. \cite{Plenior}. The parameters we used are mainly taken from Ref. \cite {TJD,t2,Safavi} which report the current experiments about the optomechanical system and OMIT. These can be used to well evaluate the feasibility.

We would also like to mention that one can also consider an atomic ensemble instead of a single atom in the system. We think that the net effect is equivalent to increasing the coupling between the single atom and cavity if the atomic ensemble is considered in the limit of large atomic number. In addition, if  two or more atoms are trapped in different even-labeled cavities, respectively, we think  multiple extra windows will occur. If they are trapped in different odd-labeled cavities, the OMIT window will change much greater . In addition, the entanglement in this optomechanical system is an interesting topic. Our preliminary results have shown the entanglement can be produced between the different components of this optomechanical system (such as between two cavities, or between one cavity and the movable mirror). It is interesting that the entanglement between the mirror and the $N$th cavity could be enhanced by the multiple cavities, but the entanglement of other components could also be reduced. In particular, the existence of the atom could play different roles in the control of the generation of various entanglement. All the detailed results will be reported in the latter papers.

In summary, we have theoretically discussed the response of an
optomechanical system which includes $N$ cavities. We have given a general
analytical expression of the generation of multiple OMIT windows. The mechanism of OMIT could have been well understood and even one could have known that an atom or atomic ensemble could broaden the width of OMIT window. However, it is shown here
that the number of the OMIT windows directly depend upon the number of
cavities. In particular, we find that, when the atom is trapped in even-labeled cavity, the
number of the OMIT windows will be increased by one; if the atom is trapped
in the odd-labeled cavity, the only the width of the OMIT windows could be
changed. In addition, we also find that the multiple OMIT windows are only
attributed to the coupling of the multiple cavities and irrespective of the
coupling to the mechanical oscillator, because the mechanical oscillator
could produce only one additional OMIT window or change the width of the OMIT windows which depends on the even- or odd- labelled cavity that the atom is trapped in.

\section*{Methods}
In this section, we will give a brief introduction of the derivation of the equations used in the main text. Substituting $\left\langle \hat{O}(t)\right\rangle =\bar{O}+\delta {O}(t)$ given in Eq. (\ref{O}) into Eqs. (\ref{(2)}-\ref{4}), Eqs. (\ref{(2)}-\ref{4}) can be rewritten by $\bar{O}$ and $\delta{O}(t)$. Since $\delta{O}(t)$ is small and depends on time and $\bar{O}$ is independent of time. One can separate equations into one related to time  and the other irrelevant of time. The equation array irrelevant of time corresponding to $\bar{O}$  has the same form as Eqs. (\ref{(2)}-\ref{4}) except setting $\varepsilon_p=0$ and $\dot{\bar{O}}=0$. In other words, if we replace $\left\langle O\right\rangle$ in Eqs. (\ref{(2)}-\ref{4}) by $\bar{O}$ and let $\varepsilon_p=0$ and $\dot{\bar{O}}=0$, we will obtain the equations for $\bar{O}$. Our Eqs. (\ref{1ping}) and (\ref{2ping}) are solved from these equations, but for avoiding repetition,  these equations are omitted here.  The equations with time corresponding to $\delta{O}(t)$ should obviously include the term $\varepsilon _{p}e^{-i\Delta t}$. They can be directly given as follows.
\begin{equation}
\delta \dot{c}_{N}=-\left( \kappa _{N}+i\Delta _{N}\right) \delta
c_{N}-ig_{N-1}\delta c_{N-1}+\varepsilon _{p}e^{-i\Delta t},  \label{(17)}
\end{equation}%
\begin{equation}
\delta \dot{c}_{n}=-\left( \kappa _{n}+i\Delta _{n}\right) \delta
c_{n}-i\left( g_{n-1}\delta c_{n-1}+g_{n}\delta c_{n+1}\right) ,\ n\neq
1,i,N, \label{(18)}
\end{equation}%
\begin{equation}
\delta \dot{c}_{i}=-\left( \kappa _{i}+i\Delta _{i}\right) \delta
c_{i}-ig_{a}\delta \sigma _{-}-i\left( g_{i-1}\delta c_{i-1}+g_{i}\delta
c_{i+1}\right) ,~i\neq 1,N, \label{(19)}
\end{equation}%
\begin{equation}
\delta \dot{c}_{1}=-\left( \kappa _{1}+i\tilde{\Delta}_{1}\right) \delta
c_{1}-ig_{1}\delta c_{2}+iG\left( \delta b^{\ast }+\delta b\right) , \label{(20)}
\end{equation}%
\begin{equation}
\delta \dot{b}=-\left( \gamma _{m}+i\omega _{m}\right) \delta b+i\left(
G\delta c_{1}^{\ast }+G^{\ast }\delta c_{1}\right) , \label{(21)}
\end{equation}%
\begin{equation}
\delta \dot{\sigma}_{-}=-\left( \gamma _{a}+i\Delta _{a}\right) \delta
\sigma _{-}+ig_{a}\delta c_{i}\bar{\sigma}_{z},  \label{(22)}
\end{equation}
where $G=g\bar{c}_{1}$ is the effective optomechanical rate. As mentioned in the text, we consider the system in the resolved sideband  regime in order to
gain more physical insight. That is, we let $\omega _{m}\gg \kappa $ and $%
\Delta _{n}=\tilde{\Delta}_{1}=\Delta _{a}=\omega _{m}$. In such a resolved
sideband regime, the lower sideband, far off-resonance can be safely
neglected. This means that in Eq. (\ref{O}), $O_{+}\approx 0$ which is the same as \cite{t2}. Thus, Eq. (\ref{(17)}-\ref{(22)}) can be rewritten for $O_-$ as
\begin{equation}
0=-\left( \kappa _{N}+ix\right) c_{N,-}-ig_{N-1}c_{N-1,-}+\varepsilon _{p},  \label{(23)}
\end{equation}%
\begin{equation}
0=-\left( \kappa _{n}+ix\right) c_{n,-}-i\left(
g_{n-1}c_{n-1,-}+g_{n}c_{n+1,-}\right) ,\ n\neq 1,i,N,  \label{(24)}
\end{equation}%
\begin{equation}
0=-\left( \kappa _{i}+ix\right) c_{i,-}-ig_{a}\sigma _{-,-}-i\left(
g_{i-1}c_{i-1,-}+g_{i}c_{i+1,-}\right) ,\ i\neq 1,N,  \label{(25)}
\end{equation}%
\begin{equation}
0=-\left( \kappa _{1}+ix\right) c_{1,-}-ig_{1}c_{2,-}+iGb_{-},  \label{(26)}
\end{equation}%
\begin{equation}
0=-\left( \gamma _{m}+ix\right) b_{-}+iG^{\ast }c_{1,-},  \label{(27)}
\end{equation}%
\begin{equation}
0=-\left( \gamma _{a}+ix\right) \sigma _{-,-}+ig_{a}c_{i,-}\bar{\sigma}_{z},  \label{(28)}
\end{equation}
where $x=\Delta -\omega _{m}$ is again the detuning from the center line of
the sideband.

\section*{Acknowledgements}
This work was supported by the National Natural Science Foundation of China,
under Grant No.11375036, the Xinghai Scholar Cultivation Plan
and the Fundamental Research Funds for the Central Universities under Grant
No. DUT15LK35 and No. DUT15TD47. A. S. is supported by China Scholarship Council (CSC) for the
Research Fellowship.
\section*{Author contributions statement}
AS conceived the idea, AS performed the calculations, YCS
 analyzed the results and wrote the main manuscript text,  ZY and ZJ participated in the discussions. All authors reviewed the manuscript.
\section*{Additional information}
Competing financial interests: The authors declare no competing financial interests.

\begin{thebibliography}{99}

\bibitem{Aspelmeyer} Aspelmeyer, M., Kippenberg, T. J. \& Marquardt, F. Cavity optomechanics. {\it Rev. Mod. Phys.} \textbf{86}, 1391 (2014).

\bibitem{e1}  Mancini, S. \textit{et al}. Entangling macroscopic oscillators exploiting radiation pressure. {\it Phys. Rev.
Lett.} \textbf{88}, 120401 (2002).
\bibitem{e2} Vitali, D. \textit{et al}. Optomechanical entanglement between a movable mirror and a cavity field.  {\it Phys. Rev.
Lett.} \textbf{98}, 030405 (2007).

\bibitem{e3} Hartmann, M. J. \& Plenio, M. B. Steady state entanglement in the mechanical vibrations of two dielectric membranes. \textit{Phys. Rev. Lett.} \textbf{101}, 200503
(2008).
\bibitem{e4} Liao, J. Q., Wu, Q. Q. \& Nori, F. Entangling two macroscopic mechanical mirrors in a two-cavity optomechanical system. {\it Phys. Rev. A} \textbf{89}, 014302
(2014).
\bibitem{AN} Nunnenkamp, A., B\o rkje, K. \& Girvin, S. M. Single-photon optomechanics. {\it Phys. Rev. Lett.}
\textbf{107}, 063602 (2011).

\bibitem{Rabl} Rabl, P. Photon Blockade Effect in Optomechanical Systems. {\it Phys. Rev. Lett.} \textbf{107}, 063601 (2011).
\bibitem{Purdy} Purdy, T. P. \textit{et al}. Strong optomechanical squeezing of light. {\it Phys. Rev. X} \textbf{3}, 031012 (2013).

\bibitem{Tian} Tian, L. Optoelectromechanical transducer: Reversible conversion between microwave and optical photons. {\it Ann. Phys.} (Berlin) \textbf{527}, 1 (2015).

\bibitem{Chang} Chang, D. E., Ni, K. K., Painter, O. \& Kimble, H. J. Ultrahigh-Q mechanical oscillators through optical trapping. {\it New J. Phys.} \textbf{14}, 045002 (2012).

\bibitem{Akram} Akram, M. J. \& Saif, F. Adiabatic population transfer based on a double stimulated raman adiabatic passage. {\it J. Russ. Laser Res.} \textbf{35} (6), 547 (2014).

\bibitem{Clerk} Wang, Y. D. \& Clerk,  A. A. Using Interference for high fidelity quantum state transfer in optomechanics. {\it Phys. Rev. Lett.} \textbf{108}, 153603 (2012).

\bibitem{Karuza} Karuza, M., \textit{et al}. Tunable linear and quadratic optomechanical coupling for a tilted membrane within an optical cavity: theory and experiment. {\it J. Opt.} \textbf{15}, 025704 (2013).

\bibitem{Hammerer} Hammerer, K., S\o rensen, A. S.  \& Polzik, E. S. Quantum interface between light and atomic ensembles. {\it Rev. Mod. Phys.} \textbf{82}, 1041 (2010).

\bibitem{Saif} Saif, F., LeKien, F. \& Zubairy, M. S. Quantum theory of a micromaser operating on the atomic scattering from a resonant standing wave. {\it Phys. Rev. A} \textbf{64}, 043812 (2001).

\bibitem{Islam} Islam, R. U., Ikram, M. \& Saif, F. Engineering maximally entangled N-photon NOON field states using an atom interferometer based on Bragg regime cavity QED. {\it J. Phys. B: At. Mol. Opt. Phys.} \textbf{40}, 1359 (2007).


\bibitem{c1} Bhattacharya, M. \& Meystre, P. Trapping and cooling a mirror to its quantum mechanical ground state. \textit{Phys. Rev. Lett.} \textbf{99}, 073601
(2007).
\bibitem{TJD} Teufel, J. D. \textit{et al}. Sideband cooling of micromechanical motion to the quantum ground state. {\it Nature} \textbf{475}, 359 (2011).

\bibitem{c3} Chan, J. \textit{et al}.   Laser cooling of a nanomechanical oscillator into its quantum ground state. {\it Nature} (London) \textbf{478}, 89 (2011).

\bibitem{Meystre} Meystre, P. A short walk through quantum optomechanics. {\it Ann. Phys.} (Berlin) \textbf{525}, 215 (2013).

\bibitem{Connell} O'Connell, A. D. \textit{et al}. Quantum ground state and single-phonon control of a mechanical resonator. {\it Nature} \textbf{464}, 697 (2010).

\bibitem{Kiesel} Akram, U., Kiesel, N., Aspelmeyer, M. \& Milburn,  G. J. Single-photon opto-mechanics in the strong coupling regime. {\it New J. Phys.} \textbf{12}, 083030 (2010).

\bibitem{Rogers} Rogers, B., \textit{et al}. Hybrid optomechanics for quantum technologies. {\it Quantum Measurements and Quantum Metrology} \textbf{2}, 11 (2014).

\bibitem{Kippenberg} Kippenberg, T. J. \& Vahala, K. J. Cavity optomechanics: Back-Action at the Mesoscale. {\it Science} \textbf{321}, 1172 (2008).

\bibitem{Restrepo} Restrepo, J., Ciuti, C. \& Favero,  I. Single-polariton optomechanics. {\it Phys. Rev. Lett.} \textbf{112}, 013601 (2014).

\bibitem{Agarwal} Agarwal, G. S. \& Huang, S. Electromagnetically induced transparency in mechanical effects of light. {\it Phys. Rev. A} \textbf{81}, 041803 (2010).

\bibitem{SM}  Huang, S. \& Agarwal,  G. S. Electromagnetically induced transparency with quantized fields in optocavity mechanics. {\it Phys. Rev. A} \textbf{83}, 043826 (2010).

\bibitem{t2} Weis, S. \textit{et al}. Optomechanically induced transparency. {\it Science} \textbf{330}, 1520 (2010).

\bibitem{Safavi} Safavi-Naeini, A. H., \textit{et al}. Electromagnetically induced transparency and slow light with optomechanics. {\it Nature (London)}\textbf{472}, 69 (2011)

\bibitem{t4} Dong, C. \textit{et al}.Transient optomechanically induced transparency in a silica microsphere. {\it Phys. Rev. A} \textbf{87}, 055802 (2013).

\bibitem{HX} Xiong, H. \textit{et al}.  Higher-order sidebands in optomechanically induced transparency. {\it Phys.
Rev. A} \textbf{86}, 013815 (2012).

\bibitem{MAL}  Lemonde, M. A., Didier, N. \& Clerk, A. A. Nonlinear interaction effects in a strongly driven optomechanical cavity. {\it Phys. Rev. Lett.} \textbf{111}, 053602 (2013).

\bibitem{KB} B\o rkje, K. \textit{et al}.
Signatures of Nonlinear Cavity Optomechanics in the weak Coupling regime, {\it Phys. Rev. Lett.} \textbf{111}, 053603 (2013).

\bibitem{AK} Kronwald, K. \& Marquardt, F. Optomechanically induced transparency in the nonlinear quantum regime. {\it Phys. Rev. Lett.} \textbf{111}, 133601 (2013).

\bibitem{MK} Karuza, M. \textit{et al}. Optomechanically induced transparency in a membrane-in-the-middle setup at room temperature. {\it Phys. Rev. A} \textbf{88}, 013804 (2013).

\bibitem{t6} Hou, B. P., Wei,  L. F. \& Wang, S. J. Optomechanically induced transparency and absorption in hybridized optomechanical systems, {\it Phys. Rev. A} \textbf{92}, 033829(2015).

\bibitem{Char} Zhang, J. Q. \textit{et al}. Precision measurement of electrical charge with optomechanically induced transparency, {\it Phys. Rev. A } \textbf{86}, 053806 (2012).

\bibitem{router} Agarwal, G. S. \& Huang, S. Optomechanical systems as single-photon routers. {\it Phys. Rev. A} 85, 021801 (2012).

\bibitem {at1} Pirkkalainen, J. M. \textit{et al}. Cavity optomechanics mediated by a
quantum two-level system. {\it Nature Communications} \textbf{6}, 6981 (2014).

\bibitem{KQu} Qu, K. \& Agarwal, G. S. Phonon-mediated electromagnetically induced absorption in hybrid opto-electromechanical systems. {\it Phys. Rev. A} \textbf{87}, 031802(R) (2013).

\bibitem{HW} Wang, H. \textit{et al}. Optomechanical analog of two-color electromagnetically induced transparency: Photon transmission through an optomechanical device with a two-level system. {\it Phys.
Rev. A} \textbf{90}, 023817 (2014).

\bibitem {atc1} Genes, C., Ritsch, H. \& Vitali, D. Micromechanical oscillator ground-state cooling via resonant intracavity optical gain or absorption. {\it Phys. Rev. A} \textbf{80}, 061803(R)
(2009).
\bibitem {atc2} Hammerer, K. \textit{et al}. Optical lattices with micromechanical mirrors. {\it Phys. Rev. A} \textbf{82},
021803(R) (2010).

\bibitem {atc3} Camerer, S. \textit{et al}. Realization of an optomechanical interface between ultracold atoms and a membrane. \textit{Phys. Rev. Lett.} \textbf{107}, 223001 (2011).

\bibitem {ures} Bariani, F.  \textit{et al}.  Hybrid optomechanical cooling by atomic $\Lambda$ systems. {\it Phys. Rev. A} \textbf{90}, 033838 (2014).

\bibitem{Ian} Ian, H., \textit{et al}. Cavity optomechanical coupling assisted by an atomic gas. {\it Phys. Rev. A} \textbf{78}, 013824 (2008).

\bibitem{LN} Neumeier, L., Leib, M.  \& Hartmann,  M. J. Single-photon transistor in circuit quantum electrodynamics. {\it Phys. Rev. Lett.} \textbf{111}, 063601 (2013).

\bibitem{CHY} Yan, C. H., Jia, W. Z. \&  Wei, L. F. Controlling single-photon transport with three-level quantum dots in photonic crystals. {\it Phys. Rev. A} \textbf{89}, 033819 (2014).

\bibitem{LZ} Zhou, L., Yang, L. P., Li, Y. \& Sun,  C. P. Quantum routing of single photons with a cyclic three-Level system. {\it Phys. Rev. Lett.} \textbf{111}, 103604 (2013).

\bibitem{JL} Lu, J., Zhou, L., Kuang, L. M. \& Nori,  F. Single-photon router: coherent control of multichannel scattering for single photons with quantum interferences. {\it Phys. Rev. A} \textbf{89}, 013805 (2014).

\bibitem{Plenior} Hartmann, M. J., Brandal, F. G. S. L.  \& Plenio,  M. B. Quantum many-body phenomena in coupled cavity arrays {\it Laser \& photon. Rev.} \textbf{2}, 527 (2008).

\bibitem{Optics} D. F. Walls and G. J. Milburn, \textit{Quantum Optics } CH 7, 128-131 (Springer, Berlin, 1994).

\bibitem{Lin} Lin, Q., \textit{et al}. Coherent mixing of mechanical excitations in nano-optomechanical structures. {\it Nat. Photonics} \textbf{4}, 236 (2010).

\bibitem{Zheng} Zheng, C., \textit{et al}. Controllable optical analog to electromagnetically induced transparency in coupled high-Q microtoroid cavities. {\it Opt. Express.} \textbf{20}, 18319 (2012).

\bibitem{Astafiev} Astafiev, O., \textit{et al}.
 Single artificial-atom lasing. {\it Nature} \textbf{449}, 588 (2007).
 
 \bibitem{SMM}Huang, S. \& Tsang, M. Electromagnetically induced transparency and optical memories in an optomechanical
system with N membranes. {\it arXiv:}1403.1340v1.

 \bibitem{Tar} Tarhan, D., Huang, S. \& M\"ustecaplio\u{g}lu, \"{O}. E. Superluminal and ultraslow light propagation in optomechanical systems. 
{\it Phys. Rev. A} \textbf{87}, 013824 (2013).


\end{thebibliography}
\end{document}